\pdfoutput=1

\documentclass[onecolumn,showpacs,preprintnumbers,amsmath,amssymb]{revtex4}

\usepackage{graphicx}
\usepackage{dcolumn}
\usepackage{bm}

\begin{document}

\preprint{FENS 07}

\title{  Cluster Expansion Method for  Evolving Weighted Networks Having Vector-like Nodes}

\author{M. Ausloos}
\email{marcel.ausloos@ulg.ac.be}

\affiliation{
GRAPES, 
SUPRATECS, U.Lg, B5a Sart-Tilman, B-4000 Li\`ege, Belgium, Euroland
}

\author{M. Gligor}
\email{mrgligor@yahoo.com}

\affiliation{National College ÒRoman VodaÓ Roman-5550, Neamt, Romania}

\date{27/02/2008}

\begin{abstract}
The Cluster Variation Method known in statistical mechanics and condensed matter is revived for weighted bipartite networks. The decomposition of a Hamiltonian through a finite number of components, whence serving to define variable clusters, is recalled. As an illustration the network  built from data representing correlations between (4) macro-economic features, i.e.  the so called $vector$ $components$, of 15 EU countries, as (function) nodes, is discussed. We show that statistical physics principles, like the maximum entropy criterion points to clusters, here in a (4) variable phase space:  Gross Domestic Product (GDP),  Final Consumption Expenditure (FCE),   Gross Capital Formation (GCF) and   Net  Exports (NEX).   It is observed that  the $maximum$ entropy  corresponds to a cluster which does $not$ explicitly include  the  GDP but only the other (3) ''axes'', i.e.  consumption, investment and trade components. On the other hand,   the $minimal$ entropy  clustering scheme is obtained from a coupling necessarily including GDP and   FCE. The results confirm intuitive economic theory and practice expectations at least as regards geographical connexions. The technique can of course be applied to many other cases in the physics of socio-economy networks.
\end{abstract}


\maketitle
\section{Introduction}

In physics one is often interested about models with  a finite number $N$ of degrees of freedom, hereby denoted by $\vec{ s} = {(s_1, s_2,\dots É, s_N)}$, taking sometimes $discrete$ values, in contrast to continuous  ones, as in field theories. For instance, the variables $s_i$ could take values $ [0$ or $1] $ (binary variables), $[-1, +1] $ (Ising spins), or $ [1, 2, . . . q]$ (Potts variables).     Network nodes and/or links can possess such degrees of freedom which indicate the role of  a few variables for characterizing or tying nodes together; these variables serve, e.g., to be exemplifying clusters, communities, ... in the network. Several network characterization techniques based on related discrete value algebra exist in the literature \cite{albert,dorogovtsev}.

Recall that statistical mechanical models are defined through an energy function, like a Hamiltonian, $\cal H $=$\cal H$$(s)$;  the corresponding probability distribution at thermal equilibrium is the Boltzmann distribution:
     
     \begin{equation}      \label{1}
     p(s) = \frac{1}{\cal Z} exp[-{\cal H} (s)]
     \end{equation}
     
     where the inverse temperature $ \beta = (k_BT )^{-1}$ 1 has been absorbed into the Hamiltonian as often conventionally done;
      \begin{equation}       \label{2}
{\cal Z}=  exp[-{\cal F}] = \sum_s  exp[-{\cal H}(s)]
     \end{equation}
     is called  the partition function and ${\cal F}$    the free energy. The Hamiltonian is typically a sum of terms, each involving a small number of variables. 
                    
 A technique which has been of interest a long time ago in condensed matter is the cluster variation approximation method \cite{CVAM,CVAMKKW,CVAMSGB}. The free energy or the Hamiltonian is expanded through a series in the variables by a systematic projection in order to define the interaction energy at each successive  cluster size level. We re-introduce the technique here, suggesting its power for discussing network properties. We take as an example and for illustration a finite size network,  one made of nodes being EU countries characterized by their most usual (macroeconomic) features. The fluctuation correlations between these serve to define the so called adjacency matrix, whence the $weights$ of the links of the network. 
  
 The technique appears to be very general and could be useful to sort out features not observed otherwise.

\section{Theoretical considerations}
   
          A useful representation is given by the {\it factor graph}. A factor graph \cite{pelizzola}  is a bipartite graph made of variable nodes $i, j, ... $, one for each variable, and function nodes $a, b, . . .$, one for each term of the Hamiltonian. A link  joins a variable node  $i$ and a function node $a$ if and only if $i \in a$, that is the variable $s_i$ appears in $H_a$, $the$ term of the Hamiltonian associated to $a$. The Hamiltonian can then be written as
                \begin{equation}      \label{3}
{\cal H} =   \sum_{a}^{N}   {\cal H}_a(s_a)
     \end{equation}
          with $s_a \equiv\{s_i, i \in a\}$  ..... This sort of writing through the decomposition of a Hamiltonian into terms describing clusters of  different (increasing) sizes has been shown to be of great interest, see \cite{clusterRNG} when, e.g. applying techniques like the renormalization group.

\begin{table*}
\caption{The minimal path length (MPL) distances to the ''average country''. Indicator: FCE ($\equiv s_2$). The moving time window size is $T$= 5 years for the data \cite{wb} taken from 1994 to 2003.}
\bigskip
\begin{ruledtabular}
\begin{footnotesize}
\begin{center}
\begin{tabular}{c c c c c c c c c c c c c c c c}

 & AUT & BEL & DEU	& DNK	& ESP	& FIN	& FRA	& GBR	& GRC	& IRL	& ITA	& LUX	& NLD	& PRT	& SWE
\\\hline 
\\94-98 & 0.88 & 0.65 &	0.85 &	0.88 &	0.65 &	0.37 &	0.65 &	0.65 &	0.65 &	0.65 &	0.37 &	0.65 &	0.65 &	0.65 &	0.65\\
\\95-99 &  0.79 &	0.79 &	0.79 &	0.81 &	0.79 &	0.41 &	0.79 &	0.79 &	0.93 &	0.79 &	0.53 &	0.59 &	0.79 &	0.79 &	0.79\\
\\ 96-00 &	1.02 &	1.02 &	1.02 &	1.02 &	1.02 &	1.02 &	1.02 &	1.02 &	1.02 &	1.02 &	0.26 &	1.02 &	1.02 &	1.02 &	1.02 \\
\\ 97-01 &	0.51 &	0.51 &	0.51 &	0.65 &	0.51 &	0.73 &	0.88 &	0.51 &	0.65 &	0.51 &	0.33 &	0.88 &	0.51 &	0.51 &	0.51 \\
\\ 98-02 &	0.52 &	0.52 &	0.52 &	0.96 &	0.52 &	0.66 &	0.95 &	0.65 &	0.96 &	0.52 &	0.35 &	1.19 &	0.52 &	0.52  & 0.52 \\
\\ 99-03 &	0.45 &	0.42 &	0.45 &	1.00 &	0.45 &	0.53 &	0.40 &	0.46 &	1.00 &	0.42 &	0.30 &	0.92 &	0.45 & 0.45 &	0.45 \\
\\\hline

\end{tabular}
\end{center}
\end{footnotesize}
\end{ruledtabular}
\end{table*}

\begin{table*}
\caption{The correlation matrix of EU-15 country movements inside the hierarchy. Indicator: FCE. The moving time window size is 5 years for data taken from 1994 to 2003.}
\bigskip
\begin{ruledtabular}
\begin{footnotesize}
\begin{center}
\begin{tabular}{c c c c c c c c c c c c c c c c}

 & $\bf{AUT}$ & $\bf{BEL}$ & $\bf{DEU}$	& DNK	& $\bf{ESP}$	& FIN	& FRA	& $\bf{GBR}$	& GRC	& $\bf{IRL}$	& ITA	& LUX	& $\bf{NLD}$	& $\bf{PRT}$	& $\bf{SWE}$
\\\hline 
\\$\bf{AUT}$ & 1.00 & $\bf{0.92}$ &	$\bf{1.00}$ & 0.23 & $\bf{0.92}$ &	0.21 & 0.38  & 0.87 &	0.03 &	$\bf{0.92}$ &	0.07 &	-0.34 &	$\bf{0.92}$ &	$\bf{0.92}$ &	$\bf{0.92}$ \\
\\$\bf{BEL}$ & &  1.00 &	$\bf{0.94}$ &	0.23 &	$\bf{1.00}$ &	0.45 &	0.56 &	$\bf{0.97}$ &	0.28 &	$\bf{1.00}$ &	0.06  &	-0.15 &	$\bf{1.00}$ &	$\bf{1.00}$ &	$\bf{1.00}$\\
\\$\bf{DEU}$ & &	& 1.00 &	0.24 &	$\bf{0.93}$ &	0.24 &	0.40 &	0.89 &	0.07 &	$\bf{0.94}$ &	0.07 &	-0.32 &	$\bf{0.93}$ &	$\bf{0.93}$ &	$\bf{0.93}$\\
\\DNK & & &	& 1.00 &	0.26 &	0.22 &	-0.14 &	0.35 &	0.75 &	0.23 &	-0.41 &	0.44 &	0.26 &	0.26 &	0.26\\
\\$\bf{ESP}$ & & & & &	1.00 &	0.45 &	0.53 &	0.97 &	0.31 &	$\bf{1.00}$ &	0.04 &	-0.15 & $\bf{1.00}$ &	$\bf{1.00}$ &	$\bf{1.00}$\\
\\FIN &	& & & & & 1.00 &	0.65 &	0.49 &	0.34 &	0.45 &	-0.68 &	0.68 &	0.45 &	0.45 &	0.45\\
\\FRA &	& & & & & & 1.00 &	0.64 &	0.05 &	0.56 &	-0.05 &	0.38 &	0.53 &	0.53 &	0.53\\
\\$\bf{GBR}$ & & & & & & & & 1.00 & 0.40 & $\bf{0.97}$ &	0.03 &	0.02 &	$\bf{0.97}$ &	$\bf{0.97}$ &	$\bf{0.97}$ \\
\\GRC & & &	& & & & & & 1.00 &	0.28 &	-0.11 &	0.45 &	0.31 &	0.31 &	0.31\\
\\$\bf{IRL}$ &	& & & & & & & & & 1.00 &	0.06 &	-0.15 &	$\bf{1.00}$ &	$\bf{1.00}$ &	$\bf{1.00}$\\
\\ITA &	& & & & & & & & & & 1.00 &	-0.68 &	0.04 &	0.04 &	0.04\\
\\LUX &	& & & & & & & & & & & 1.00 &	-0.15 &	-0.15 &	-0.15\\
\\$\bf{NLD}$ &	& & & & & & & & & & & & 1.00 &	$\bf{1.00}$ &	$\bf{1.00}$\\
\\$\bf{PRT}$ &	& & & & & & & & & & & & &	1.00 &	$\bf{1.00}$\\
\\$\bf{SWE}$ &	& & & & & & & & & & & & & &	1.00\\
\\\hline

\end{tabular}
\end{center}
\end{footnotesize}
\end{ruledtabular}
\end{table*}

     In combinatorial optimization problems, the Hamiltonian plays the role of a {\it cost function} and one is often interested in the low temperature limit $T\rightarrow 0$, where only minimal energy states (ground states) have a nonÐvanishing probability.
    
      Probabilistic graphical models are usually defined in a slightly different way\cite{smyth}. E.g., in the case of {\it Markov random fields}, also called {\it Markov networks}, the joint distribution over all variables is given by
                     \begin{equation}      \label{4}
p(s)=    \frac{1}{{\widehat{\cal Z}}} \prod_a   \psi_a(s_a)
     \end{equation}
where $\psi_a$ is called the potential, and 
 \begin{equation}       \label{5}
 {\widehat{\cal Z}}= \sum_s  \prod_a   \psi_a(s_a).
     \end{equation}
     
          Of course, a statistical mechanical model described by the Hamiltonian (3), corresponds to a probabilistic graphical model with potentials $\psi_a = exp(-{\cal H}_a)$, and corresponding ${\cal Z} = {\widehat{\cal Z}}$
and $\cal{F}=  \widehat{\cal F}$.

Next we define a cluster $\alpha$ as a subset of the factor graph such that if a function node belongs to $\alpha$, then all the variable nodes $s_{\alpha}$ also belong to $\alpha$ ; notice that  the converse needs not to be true, otherwise the only legitimate clusters would be the connected components of the factor graph. Given a cluster we can write its probability distribution, defined as the ratio between the number of realized connections and the number of all possible connections,  as
           \begin{equation}       \label{6}
p_{\alpha}(s_{\alpha})= \sum_{s\in\alpha}  p(s).
     \end{equation}
     and its entropy
       \begin{equation}       \label{7}
    {\cal S}_{\alpha}(s_{\alpha})= - \sum_{s\in\alpha} p(s)\cdot\ln{p(s)}    
    \end{equation}

     \section{Illustration}

    As a short illustration, consider the function nodes to be countries and the variables  to be  macroeconomic indicators \cite{durlauf}, i.e.
    
    \begin{enumerate} 
     \item Consider the  nodes to be the  first (in time) 15 EU countries. Let the country names be abbreviated according to The Roots Web Surname List (RSL) \cite{codes} which uses 3 letters standardized abbreviations. 
   \item Suppose that we are interested in a vector describing each country (Hamiltonian or) ''thermodynamic state'' with 4 components,  i.e.  $s_1$ $\equiv$  Gross Domestic Product (GDP), $s_2$ $\equiv$  Final Consumption Expenditure (FCE), $s_3$ $\equiv$  Gross Capital Formation (GCF) and $s_4$ $\equiv$  Net  Exports (NEX).  The World Bank database \cite{wb} is here used as data source.  Let the data be taken from 1994 to 2004 for GDP and from 1994 to 2003 for FCE, GCF and  NEX, respectively.  
\end{enumerate}

   The yearly fluctuations of these four variables are easily  calculated and their auto- and cross-correlation matrices easily obtained; see e.g. a discussion for GDP in \cite{r1,r2} and more detail elsewhere \cite{r3}. Essentially, the correlations can calculated for a  time window of given size moving along the time axis; these are used for getting the statistical distances among countries, e.g. $A$ and $B$, 
    for various time window sizes $T$ at various times $t$, where $t$ is the final point of the interval, i.e.
 
\begin{equation}
d_s(A,B)_{(t,T)} = \sqrt{2 (1- C_{(t,T)} (A,B))} 
\label{eq:stat}
\end{equation}
where

 \begin{equation}
 C_{(t,T)}(A,B) = 
\frac{\langle AB \rangle_{(t,T)} - \langle A \rangle_{(t,T)} \langle B \rangle_{(t,T)} }{\sqrt{ ( \langle A^2 \rangle_{(t,T)} - \langle A \rangle_{(t,T)}^2)( \langle B^2 \rangle_{(t,T)} - \langle B \rangle_{(t,T)}^2)}}.
 \end{equation}
The brackets $\langle ... \rangle$ denotes the expectation value of the ''$A,B$ time series'', - here in the interval $(t-T,t)$.

These distances are thus mapped onto ultrametrical distances, as in the classical Minimum Spaning Tree (MST) method. By calculating the statistical distances {\it with respect to the average value  of the index} (seen here as for an ''average'' country), we get a country hierarchy that proves to be changing from a time interval to another when the (constant size) time window is moved over the full time span. The correlation coefficients refer to the movement of the countries inside this hierarchy.   
   
In order to exemplify this method, the corresponding steps for $s_2$ $\equiv$  FCE are explicitly shown below (for $s_1$ $\equiv$  GDP the first steps are explicitly described in \cite{r4}). After the (virtual) "average" country is introduced in the system, the statistical distances corresponding to the fixed 5 years moving time window can be calculated and set in increasing order. The minimal  path  length    (MPL) connections to the ''average'' country can be established for each country in every time interval (Table I).   
The resulting hierarchy is readily found to be changing from a time interval to another. 
The above procedure is repeated for each macroeconomic indicator, leading to similar three Tables to Table I.

Next, a time independent  ''correlation matrix'' can be built, at this stage for the {\it country  movement fluctuations inside the hierarchy}, i.e. averaging the relative MLP fluctuations between countries. In so doing, the  moving-average-minimal-path- length  (MAMPL)  method leads us to a set of $M$ = 4 correlation matrices (one for each index, 
having the size $N$ x $N$, where $N$ = 15 is the number of countries under consideration here. E.g. Table II, for FCE ($\equiv s_2$). N.B. the matrix is symmetric (half of the elements are shown)  but not all elements are necessarily positive.

 \begin{figure}
\centering
\includegraphics[height=8cm,width=8cm]{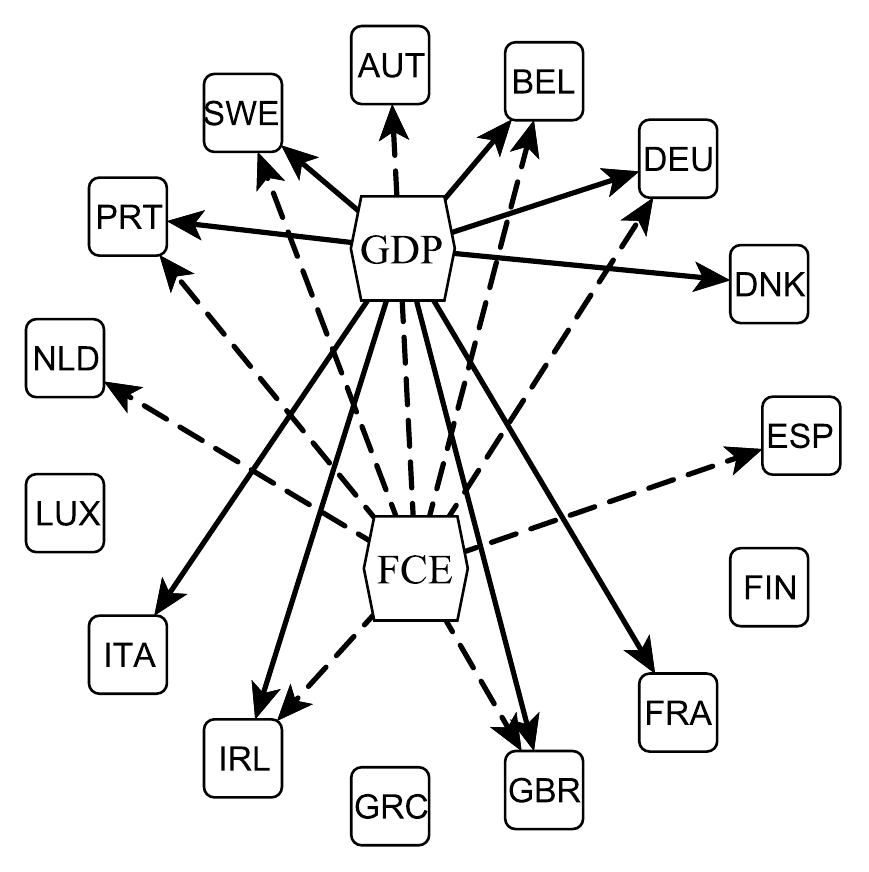}
\caption{\label{Fig1a} The factor graph associated to the first 15 EU country connections, according to the strongest correlations in the Gross Domestic Product (GDP) and Final Consumption Expenditure (FCE)}
\end{figure}

\begin{figure}
\centering
\includegraphics[height=8cm,width=8cm]{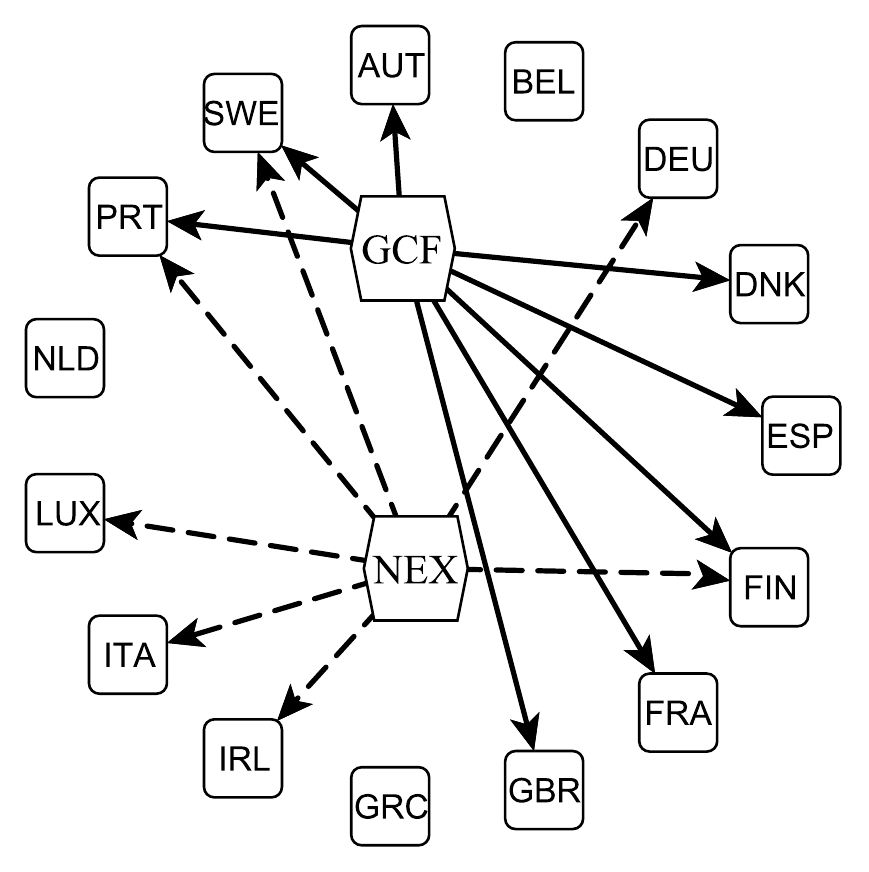}
\caption{\label{Fig1b} The factor graph associated to the first 15 EU country connections, according to the strongest correlations in  the Gross Capital Formation (GCF) and Net Exports (NEX)}
\end{figure}

Let us suppose that we filter these (four) second correlation matrices in order to retain a few  terms, - those which lead us to build
  a network for which the weights (i.e. the correlation coefficients) are greater than e.g. 0.9. The  correlation coefficients,  e.g. in the case of the variable node $s_2$ (FCE fluctuations),  as given   in Table II,  are emphasized in bold for those $\ge 0.9$ at each couple of  function nodes. 
Due to the filtering, one can easily see that not all 15 countries have at least one ''bold node'', i.e. are connected through the variable node $s_2$ (FCE), but only nine of them, namely AUT, BEL, DEU, ESP, GBR, IRL, NLD, PRT and SWE. In Fig. 1, these nine countries are connected through the variable node FCE (the dashed arrows). 

The above procedure can be repeated for GDP, GCF and NEX, whence obtaining the other ''clusters'' in Figs. 1-2, with respectively 9, 8 and 7 countries (for this filter value). Notice that GRC does not belong to any cluster \cite{GRC}. 

The cluster contributions to the Hamiltonian can thus be the variable $s_2$.    Then one obtains that the cost function $\cal H$ associated to the factor graph (Figs. 1, 2) based on these four variables reads

$\cal{H}$ =  $\cal{H}$$_1$ + $ \cal{H}$$_2$ + $  \cal{H}$$_3$  +$  \cal{H}$$_4$,

where
 
${\cal H}_1$=  (LUX)($s_4$) + (NLD)($s_2$),

${\cal H}_2$ = (ITA)($s_1, s_4$) +  (AUT)($s_2, s_3$) + (BEL)($s_1, s_2$) + (DNK)($s_1, s_3$),

${\cal H}_3$ = (ESP)($s_2, s_3$) +  (FIN)($s_3, s_4$) + 
+ (FRA)($s_1, s_3$) 
+ (DEU)($s_1, s_2, s_4$) + (GBR)($s_1, s_2, s_3$)  + (IRL)($s_1, s_2, s_4$),  

${\cal H}_4$=  (PRT)($s_1, s_2, s_3, s_4$) + (SWE)($s_1, s_2, s_3, s_4$),  

from which one could write the equilibrium probability distribution, the partition function and the free energy, introduced here above.

\section{Conclusion}

    Instead of writing a  Hamiltonian as a function of the function nodes, let us project the dynamics of the factor graph  into a  phase space spanned by the variable nodes. Recall that a cluster  $\alpha$ was defined as a subset of the factor graph such that if a function node belongs to $\alpha$, then all the variable nodes $s_{\alpha}$ also belong to $\alpha$. We can write all the possible combinations of the four variable nodes and find  the Hamiltonian  corresponding to function nodes. Let us take for example the combination ($s_{1}$ $\equiv$ GDP; $s_{2}$ $\equiv$  FCE; $s_{3}$ $\equiv$ GCF). Then, the function nodes connected $only$ to these $three$ variables (not necessarily to all of them) $and$ not to the fourth one ($s_4$ $\equiv$ NEX) are AUT, BEL, DNK, ESP, FRA, GBR and NLD. This means a cluster that we can see in the first row in Table III. The same can be done for the other three combinations, leading to another set of clusters.
     
    In so doing clustering \cite{clustering} properties appear through e.g. an entropy, Eq. (7). The values are given in Table III for the  clusters made of three variable nodes. As a not obvious consequence of this cluster analysis  technique, it is observed that  the $maximum$ entropy (0.367) corresponds to the clustering  which does $not$ explicitly include  the GDP but only the consumption, investment and trade components. Another point can be deduced from   the $minimal$ entropy (0.347) clustering scheme, i.e. it is obtained from the coupling between GDP and  FCE. The results confirm intuitive economic theory and practice expectations at least as regards geographical connexions. However deep discussions of these findings  are left for economists.
               
          In conclusion, let us recall the frame of our work and our findings : relevant microscopic description of a system
relies on a coarse-grained reduction of its internal variables. We have presented a way to do so for a bipartite graph having on one hand countries, on the other hand macro-economy indicators. We have obtained a Hamiltonian description. The technique can of course be generalized and applied to many other  socio-economy networks.

\begin{table}
\caption{Clustering of the first 15 EU countries in a 4-variable factor graph approach after filtering (see text)  and projecting in a three variable node phase space; the number of links in the cluster,  the  maximum possible number of links, subsequently the relevant ratio, and the entropy of each cluster are given}
\begin{center} \begin{tabular}{|l|r|r|r|r|r|} 
\hline variable & cluster of &  number& maximum & & \\
  & function & of & number & ratio & entropy\\
	     nodes &  nodes &  links & of links            &          & \\
  \hline 
GDP-&  -AUT-BEL- &  & &    &  \\ 
-FCE&  -DNK-ESP-FRA- &14   & 28&    0.500   & 0.347  \\ 
-GCF&  -GBR-NLD-    &   &  &        &  \\ 
\hline
GDP-& -BEL-DEU-&  &  &  &\\ 
-FCE-& -IRL-ITA-& 12& 24& 0.500 &0.347\\ 
-NEX &-LUX-NLD  &   &  &        &  \\ 
\hline
GDP-& -DNK-FIN-& & &  & \\
-GCF-& -FRA- &9 &20 &0.450 &0.359\\
-NEX &  -ITA-LUX  &   &  &        &  \\ 
\hline
FCE- & -AUT-ESP- & & & & \\ 
 -GCF-&  -FIN-  &8 &20 &0.400 &0.367\\ 
-NEX &-LUX-NLD&   &  &        &  \\ 
 
\hline \end{tabular} \end{center} \label{table1} \end{table}

 \section{Conclusion}

Complex networks have become an active field  of research in physics  \cite{pastor}.
These systems are usually composed of a large number of internal components (the nodes and links), and describe a wide variety 
of systems of high  intellectual  and technological importance. Relevant questions pertain to the characterization of the networks.  Investigations of the case of  directed and/or weighted networks are not so common. The occurrence of  community clustering for networks having nodes possessing a vector-like characteristics has been rarely studied.  We have attempted to do so through a revival of some clustering variation method in the framework of some macro-economy study.

   
   We have taken as an example the weighted fully connected network of the $N$ = 15 first countries forming the European Union in 2005 (EU-25). The ties between countries are supposed to result (be proportional) to the degree of similitude of the macroeconomic fluctuations annual rates of  $four$ macro-economic indicators, i.e. - Gross Domestic Product (GDP), Final Consumption Expenditure (FCE), Gross Capital Formation (GCF) and Net  Exports (NEX) over $ca.$ a 15 year time span.

Averaging the yearly increment correlations  a weighted bipartite network has been built having the four ''degrees of freedom'' and the fifteen ''countries'' on the other hand as basis.
The analysis  shows the importance of $N$-body
interactions  in particular when observing the   macro-economy  states   asa function of time. This   leads to identify and display clusters of countries, -clusters resulting from  projections onto a  high-dimensional phase space spanned by  indicators, taken as independent variables.  This approach generalizes
usual projection methods by accounting for the complex geometrical
 connections resulting from vector-like nodes.

 In particular such a measure of collective habits does fit the usual and practical expectations defined by politicians, journalists,  or economists,  through so called ''common factors'' \cite{mora,barrios}. The analysis reveals geographical connexions indeed.   It is expected that the technique can be applied to many types of physical and socio-economic networks.
  
\begin{acknowledgments} MG would like to thank the Francqui Foundation for financial support thereby having facilitated his stay in Li\`ege.  This work has been part of investigations stimulated and supported by the European Commission through the Critical Events in Evolving Networks (CREEN) Project (FP6-2003-NEST-Path-012864). MA thanks the FENS 07 organizers for  their invitation and as usual very warm welcome in Wroc{\l}aw.
\end{acknowledgments}


\begin{thebibliography}{0}
\bibitem{GRC}  GRC does not appear in the Hamiltonian, because (see Figs.1, 2) GRC is $not$ connected to the other countries  through any $s_i$ by a link having a weight greater than 0.9.   If the linkage threshold is established to a lower value, e.g.$ |C| \geq 0.8$, its function node appears as (GRC)($s_1, s_4$), i.e. it belongs to the same clusters as Italy.
\end{thebibliography}
\end{document}